\newif\ifabstract
\abstracttrue
 \abstractfalse 
\newif\iffull
\ifabstract \fullfalse \else \fulltrue \fi

\documentclass[11pt]{article}

\usepackage{rotating}
\usepackage{multirow}
\usepackage{amsfonts}
\usepackage{amssymb}
\usepackage{amstext}
\usepackage{amsmath}
\usepackage{xspace}
\usepackage{theorem}
\usepackage{graphicx}
\usepackage{url}
\usepackage{graphics}
\usepackage{colordvi}
\usepackage{colordvi}
\usepackage{subfigure}

\advance \oddsidemargin by -0.8in
\newcommand{\myparskip}{3pt}
\parskip \myparskip

\begin{document}

\title{Geant4 Maintainability Assessed with Respect to Software Engineering References\footnote{Presented at 2016 IEEE NSS/MIC - Strasbourg, France, 29 October - 5 November, 2016}}

\author{Elisabetta Ronchieri\thanks{INFN CNAF at Bologna. Email: {\tt elisabetta.ronchieri@cnaf.infn.it}}\and Maria Grazia Pia\thanks{INFN at Genoa. Email: {\tt maria.grazia.pia@ge.infn.it}} \and Tullio Basaglia\thanks{CERN. Email: {\tt Tullio.Basaglia@cern.ch}} \and Marco Canaparo \thanks{INFN CNAF at Bologna. Email: {\tt marco.canaparo@cnaf.infn.it}}}

\begin{titlepage}
\maketitle

\thispagestyle{empty}

\begin{abstract}
We report a methodology developed to quantitatively assess the maintainability of Geant4 with respect to software engineering references. The level of maintainability is determined by combining a set of metrics values whose references are documented in literature. 
\end{abstract}

\end{titlepage}

\section{Introduction}\label{sec: intro}

This report documents what has been done so far to assess Geant4 maintainability - one of software characteristics defined in software quality standards, such as ISO/IEC 25010:2011 (former ISO/IEC 9126) \cite{iso}.

Maintainable software allows development teams to fix bugs, add new features, improve usability and increase performance. Organizations
that deal with software  in different domains, such as telecommunications, aerospace and simulations, monitor such software characteristic to maintain skills and knowledge in order to understand and make changes to their software. 

Software characteristics are measured by metrics values. We identified and assessed software metrics tools (both free and under commercial licenses) to collect a large number of measurements \cite{ronchieri1}, \cite{ronchieri2}.
As a result of this assessment, we selected Imagix4D \cite{imagix4d}: this tool measures several product metrics at different levels, such as file, class, directory, namespace, function and variable, and its vendor positively collaborates with research communities.
Metrics used in this study provide code information about size, coupling, inheritance, control-flow structuredness, cohesion, staticness.
Existing literature gives references of such metrics for different programming languages. We identified C++ metric thresholds to determine the goodness of code.

\section{Method}
\label{sec_met}

The methods used to perform this maintainability assessment is characterized by the following steps:
\begin{enumerate}
\item collecting the source code of all Geant4 \cite{agostinelli,allison} versions from 0 to the current one (10.2);
\item loading the Geant4 source code into Imagix 4D version 8.0.4 to measure a large number of metrics in order to obtain code information about 
size, coupling, inheritance, control-flow structuredness, cohesion, staticness;
\item saving all the collected data at different levels of granularity, such as file, function, class, directory, variable and namespace;
\item application of statistical methods for the analysis of metric values;
\item identification of quality (goodness ranges of) references with respect to size, coupling, inheritance, control-flow structuredness, cohesion, staticness, derived from 
relevant  peer-reviewed papers, conference proceedings and technical reports \cite{ronchieri3}.
\end{enumerate}

Some of the metrics we collected are listed in Tables \ref{tab_gsize}, \ref{tab_gc} and \ref{tab_goo}.

\begin{table}[th]
\parbox{.45\linewidth}{
\scriptsize{
\caption{Some Metrics of the Size Group}
\label{tab_gsize}
\begin{center}
{
\begin{tabular}{lll}
\hline
Group & Size Metric &  Source \\
\hline
\multirow{9}{*}{File}  & Comment Ratio    &  Lorenz and Kidd\\
& Declarations in File     &  Lorenz and Kidd\\
& File Size     &  Lorentz and Kidd\\
& Functions in File	       &  Lorenz and Kidd\\
& Lines in File	      &  Lorenz and Kidd\\
& Lines of Source Code &  Lorenz and Kidd\\
& Lines of Comments    & Lorenz and Kidd\\
& Number of Statements & Lorenz and Kidd\\
& Variables in File        & Lorenz and Kidd\\
\hline
\multirow{3}{*}{Function} & Lines in Function & Lorenz and Kidd\\
&Lines of Source Code 	 & Lorenz and Kidd \\
&Variables in Function 	 & Lorenz and Kidd\\
\hline
\end{tabular}
}
\end{center}
}
}
\hfill
\parbox{.45\linewidth}{
\scriptsize{
\caption{Some Metrics of the Complexity Group}
\label{tab_gc}
\begin{center}
{
\begin{tabular}{lll}
\hline
Group & Complexity Metric & Source \\
\hline
File, & Intelligent Content  & Halstead\\
Function, & Mental Effort        & Halstead\\
Class & Program Volume 	    & Halstead\\
& Program Difficulty  & Halstead\\
\hline
File, & Average Cyclomatic Complexity  &McCabe\\
Class &Maximum Cyclomatic Complexity  & McCabe\\
&Total Cyclomatic Complexity      & McCabe\\
\hline
File & Maintainability Index 	& Welker \\
\hline
\multirow{4}{*}{Function} & McCabe Cyclomatic Complexity  &McCabe\\
&McCabe Decision Density	 & McCabe\\
&McCabe Essential Complexity   & McCabe\\
&McCabe Essential Density        &	 McCabe\\
\hline
\end{tabular}
}
\end{center}
}
}
\vfill
\parbox{.45\linewidth}{
\scriptsize{
\caption{Some Metrics of the Object-Oriented Group}
\label{tab_goo}
\begin{center}
{
\begin{tabular}{lll}
\hline
Group & Object-Oriented Metric &  Source \\
\hline
\multirow{6}{*}{Class} &  Class Cohesion (LCOM) &Chidamber and Kemerer\\
& Class Coupling	(CBO)            & Chidamber and Kemerer\\
& Depth of Inheritance (DIT)	& Chidamber and Kemerer\\
& Number of Children (NOC)    & Chidamber and Kemerer\\
& Response for Class (RFC)     & Chidamber and Kemerer\\
& Weighted Methods (WMC)	& Chidamber and Kemerer\\
\hline
\end{tabular}
}
\end{center}
}
}
\end{table}

\begin{figure}[th] 
\centering
\begin{minipage}{.45\textwidth}
\centering
\includegraphics[angle=0,width=6.1cm]{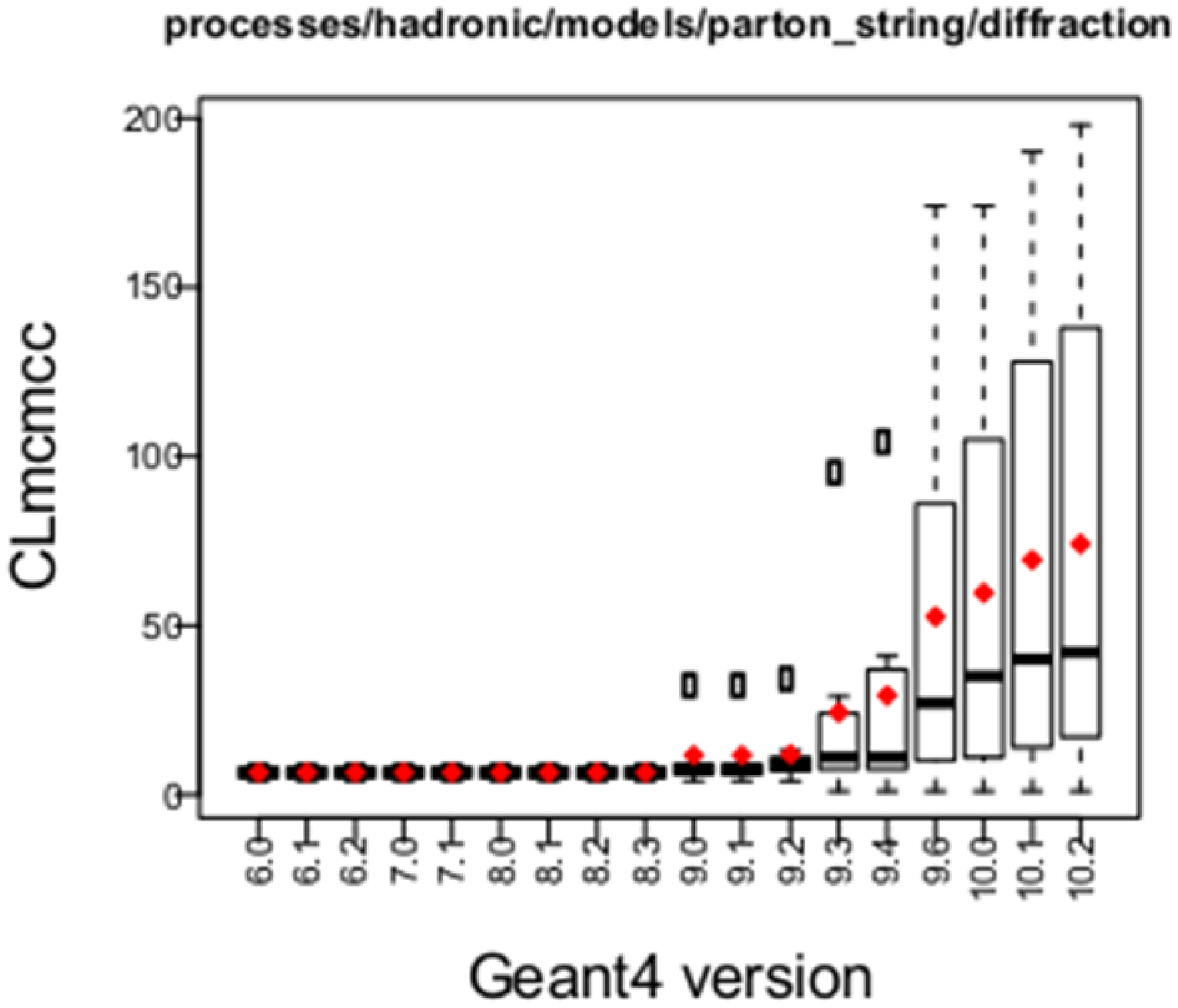}
\caption{McCabe Maximum Cyclomatic Complexity at class level for the diffraction package.}
\label{fig_mcmcc}
\end{minipage}
\hfill
\begin{minipage}{.45\textwidth}
\centering
\includegraphics[angle=0,width=6.3cm]{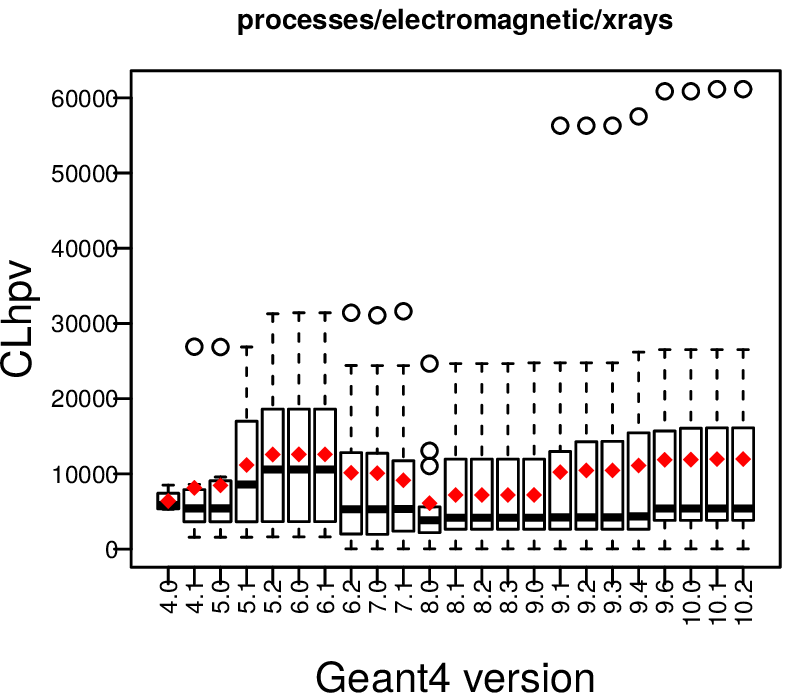}
\caption{Halstead’s programme volume at class level for the xrays package.}
\label{fig_hpv}
\end{minipage}
\end{figure}

Size metrics \cite{Lorenz} quantify code size. They are estimators of software cost and effort.

Complexity metrics, such as McCabe \cite{McCabe} and Halstead \cite{Halstead}  metrics,
measure the simplicity of the system design. 
McCabe’s complexity, also called cyclomatic complexity, 
quantifies the control flow within a program by counting the independent paths
on a control flow graph. The path indicates a certain degree of well
structuredness of an application.

Object-oriented metrics \cite{Chidamber} measure complexity, 
maintenance and clarity; they estimate to which extent the system 
adheres to the object orientation. 

\section{A sample of quality references and results}

Figure \ref{fig_mcmcc} shows the trend of McCabe Maximum Cyclomatic Complexity at class level for 
Geant4 hadronic physics \textit{diffraction} package, 
while Figure \ref{fig_hpv} shows the trend of Halstead' programme volume for the electromagnetic physics \textit{xrays} package.

Table \ref{tab_ref} shows a sample of quality references. 
\begin{table}[h]
\scriptsize{
\caption{A Sample of Quality References}
\label{tab_ref}
\begin{center}
{
\begin{tabular}{lll}
\hline
Acronym & Reference &  Source \\
\hline
Comment Ratio & 0.08 & McCabe\\
SLOC (Source Lines Of Code) & 60 at file level & McCabe\\
\hline
\multirow{6}{*}{HPV (Halstead Programme Volume)} & 1500 at function level & McCabe\\
& [100,8000] at file level & Verysoft Technology \cite{vt}\\
& $>$ 800 too many things at file level & Verysoft Technology \cite{vt}\\
& [20, 1000] at function level & Verysoft Technology \cite{vt}\\
& $>$ 1000 too many things at function level
& Verysoft Technology \cite{vt}\\
\hline
\multirow{3}{*}{MI (Maintainability Index)}& $<$65 poor maintainability & Coleman, Lowther, Oman \cite{lowther}
\\
& [65, 84] fair maintainability &Coleman, Lowther, Oman \cite{lowther}\\
& $\ge$85 excellent maintainability&Coleman, Lowther, Oman \cite{lowther}\\
\hline
\multirow{8}{*}{MCMCC (McCabe’s Maximum Cyclomatic Complexity} & [1, 10] low CC & CppDepend \cite{cpp}\\
& [11, 15] medium CC& CppDepend \cite{cpp}\\
& [16,30] high CC& CppDepend \cite{cpp}\\
& $>$31 very high CC& CppDepend \cite{cpp}\\
& [1, 10] low CC& McCabe\\
& [11, 20] medium CC& McCabe\\
& [21. 50] high CC& McCabe\\
& $>$51& McCabe\\
\hline
\end{tabular}
}
\end{center}
}
\end{table}

\section{Conclusions}\label{sec: con}

The use of metrics can contribute to monitor the internal quality of software. 
Further investigation is in progress to identify appropriate ranges of metric values for Geant4 packages by using statistical methods.
More extensive results will be discussed in a forthcoming full paper.

\section*{Acknowledgment}

The authors thank Francesco Giacomini for interesting discussions and INFN CCR for partly funding this work. We also thank the Imagix Corporation that provided an extended free full license of Imagix 4D for performing this work and CERN library for providing papers and books.

\bibliography{nss}
\bibliographystyle{alpha}

\end{document}